\documentclass[11pt]{article} \usepackage{proc}
\usepackage{amssymb} \usepackage{german} 
\usepackage{graphicx} \DeclareGraphicsExtensions{.eps}

\newcommand{\adlo}{{\sc ADLO}}
\newcommand{\aleff}{{\sc Aleph}}
\newcommand{\delphi}{{\sc Delphi}}
\newcommand{\ldrei}{{\sc L3}}
\newcommand{\opal}{{\sc Opal}}
\newcommand{\lnln}{\mbox{$l\nu\,l'\nu'$}}
\newcommand{\qqln}{\mbox{$q\bar{q}'\,l\nu$}}
\newcommand{\qqqq}{\mbox{$q\bar{q}'\,q\bar{q}'$}}
\newcommand{\gev}{\,{\rm GeV}}

\newcommand{\hftwo}{\hspace*{\fill}}

\newcommand{\mev}{\,{\rm MeV}}

\newcommand{\pb}{\,{\rm pb}}

\newcommand{\qi}{``}
\newcommand{\qo}{''}

\newcommand{\eprintk}[2]{{hep-#1/}#2,}

\newcommand{\jrnl}[4]{{#1} {\bf #2} (#3) #4.}

\newcommand{\EPJC}{{\em Eur.\ Phys.\ J.\ }{\bf C}}

\newcommand{\NPPS}{\em Nucl.~Phys.\ Proc.~Suppl.}

\begin{document}
\vspace*{4cm}
\title{W MASS AND ITS UNCERTAINTY FROM MODELLING THE
  HADRONIC FINAL STATE AT LEP}

\author{K.~RABBERTZ}

\address{EP Division, CERN, CH-1211 Geneva 23, Switzerland\\
  e-mail: klaus.rabbertz@cern.ch}

\maketitle\abstracts{From 1996 up to 2000 the LEP collider at CERN has
  operated at center of mass energies above the production threshold
  for $W$ boson pairs of approximately $160\gev$. The obtained data
  are used to extract a preliminary $W$ mass value of $(80.450 \pm
  0.039)\gev$ by direct reconstruction. To a large extent the
  uncertainty is due to systematic effects especially in the fully
  hadronic decay channel $W^+W^- \rightarrow$ \qqqq\ that suffers most
  from ambiguities in modelling the hadronic final state. Methods to
  assess and reduce these uncertainties are the current main concern
  of the four LEP experiments.}

\section{Current Status of the W Mass Measurement at LEP}

Each of the four experiments \aleff, \delphi, \ldrei\ and \opal\ has
gathered around $700\pb^{-1}$ of integrated luminosity corresponding
to about $10000\,WW$ pairs per experiment. Depending on the decay
products of the $W$ bosons the possible configurations of the final
state fall into three categories, the leptonic $W^+W^- \rightarrow$
\lnln, the semi-leptonic $W^+W^- \rightarrow$ \qqln\ and the hadronic
$W^+W^- \rightarrow$ \qqqq\ channel with branching ratios of $11\%$,
$44\%$ and $45\%$ respectively. Given reconstruction efficiencies of
$80$--$85\%$ and purities around $90\%$ ($80\%$ for the hadronic
channel) it is evident that the measurement of the $W$ mass at LEP is
dominated by semi-leptonic and hadronic $WW$ events from which it can
directly be reconstructed.  The leptonic channel, which in addition
suffers from the fact that it contains at least two undetected
neutrinos in the final state, can merely serve as a cross-check. It is
not affected, however, by uncertainties in the description of the
transition from quarks to jets of hadrons (hadronization) and may
become important in a future linear collider with very high
luminosity. Final results from an \opal\ analysis~\cite{opallnln} are
given in table~\ref{tab:dmw}.\footnote{Note that the systematic
  uncertainty contains a statistical component and would shrink with
  higher integrated luminosity.}

\begin{figure}[b]
  \hftwo\includegraphics[height=6.25cm]{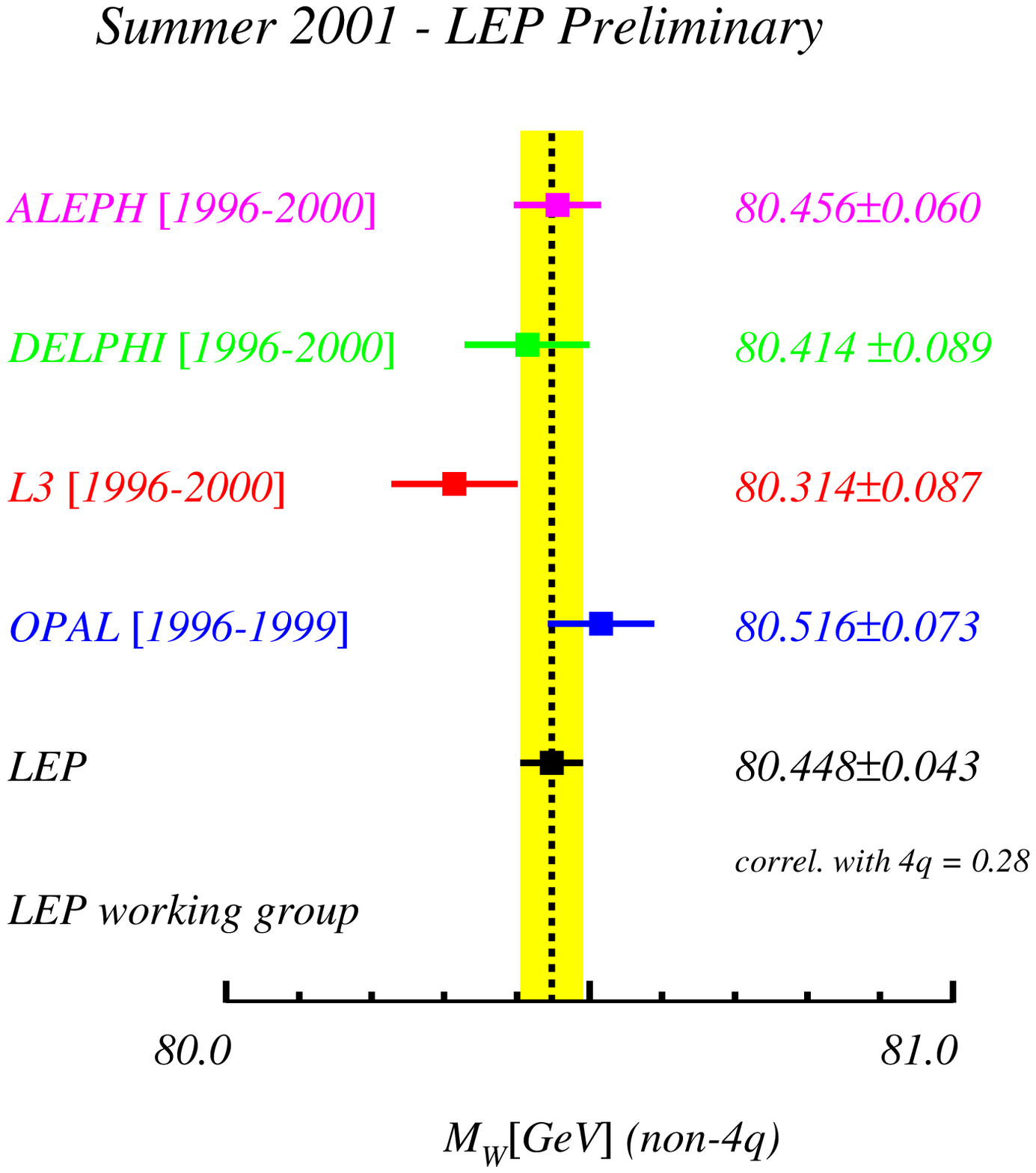}\hftwo%
  \includegraphics[height=6.25cm]{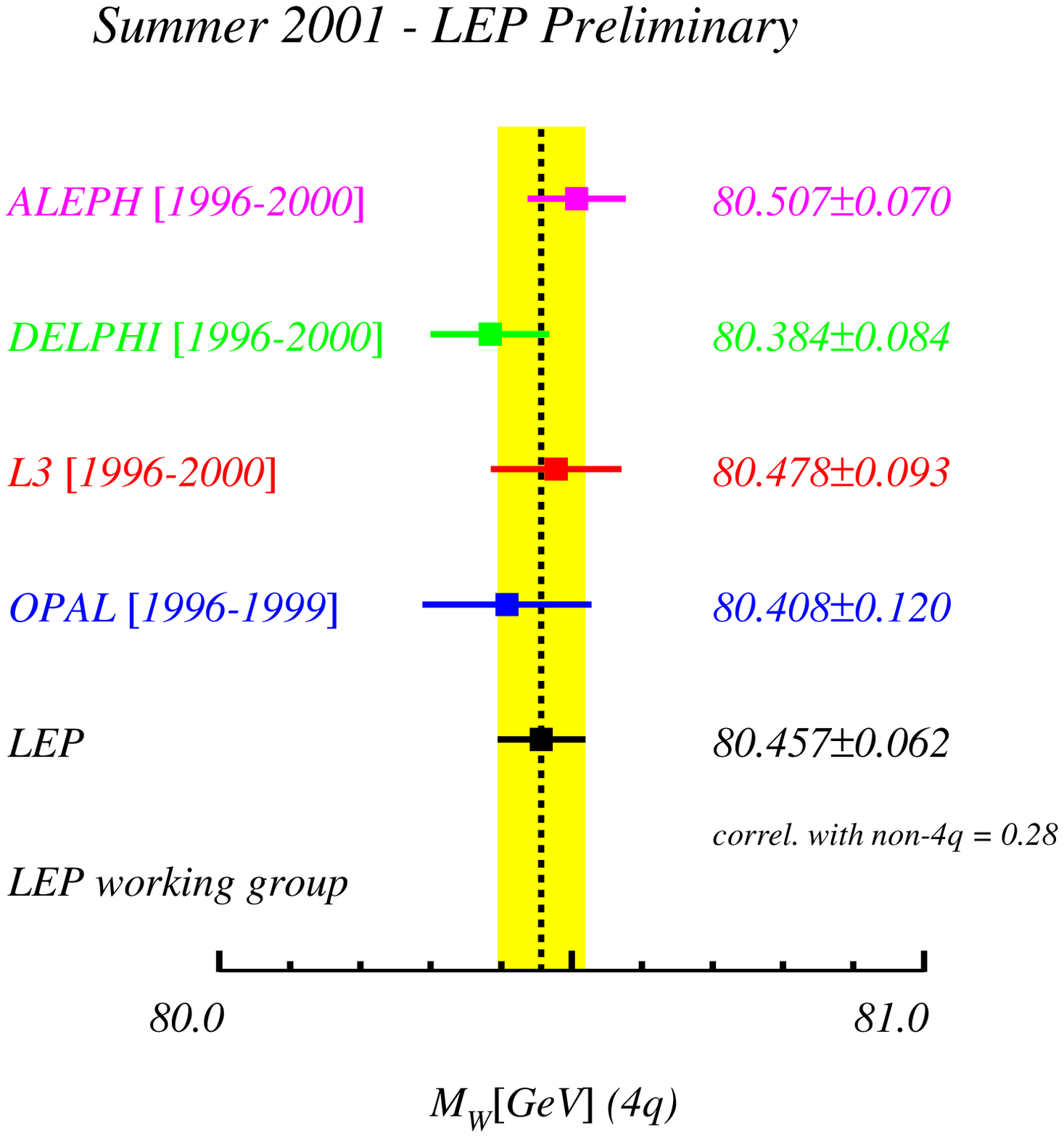}\hftwo
  \caption{Preliminary $W$ mass measurements from LEP}
  \label{fig:mw}
\end{figure}

For the hadronic $WW$ events additional complications arise with
respect to the semi-leptonic ones because of ambiguities in the
assignment of jets to $W$'s on the one hand and possible cross-talk
either between the quarks from different $W$'s before the formation of
hadrons due to colour reconnection (CR) or Bose-Einstein correlations
(BEC) between identical bosons on the other hand.  This leads to the
unpleasant fact that despite a larger statistical weight of $55\%$
compared to $45\%$ the hadronic channel contributes only to $27\%$ to
the final combination. The latest preliminary results on the $W$ mass
and its uncertainty from a combination of all four LEP experiments
\adlo~\cite{lepcomb} can be found in fig.~\ref{fig:mw} and
table~\ref{tab:dmw}. Obviously, it is worthwhile to look for a
systematic bias on $M_W$ due to additional final state interactions
possible in the hadronic channel. To date, no significant discrepancy
between $M_W^{\rm non-4q} = (80.448 \pm 0.043)\gev$ and $M_W^{\rm 4q}
= (80.457 \pm 0.062)\gev$ has been found:~\cite{lepcomb} $\Delta
M_W(\{{\rm 4q}\}-\{{\rm non-4q}\}) = (+9 \pm 44)\mev$.

In the following, the effects of colour reconnection and hadronization
will be discussed in more detail, Bose-Einstein correlations are
covered elsewhere.\cite{thomas}

\begin{table}[t]
  \caption{Decomposition of the W mass uncertainty for the final
    OPAL analysis of the leptonic decay channel and for the preliminary
    LEP combined results of the semi-leptonic and hadronic channels.
    Detector systematics include uncertainties in the jet and lepton
    energy scales and resolution. The \qi{}Other\qo{} category refers to
    sources of uncertainty largely uncorrelated between the experiments
    like event selection, fitting method, background estimation,
    simulation statistics or four-fermion treatment.}
  \begin{center}
    \begin{tabular}{|l|r||r|r|r|}
      \hline & \multicolumn{4}{|c|}{$\Delta M_W/{\rm MeV}$}\\\hline
      & O final & \multicolumn{3}{c|}{LEP ADLO Preliminary}\\\cline{2-5}
      \raisebox{1.0ex}[-1.5ex]{Data} &
      \multicolumn{1}{c||}{\lnln} & \qqln & \qqqq & Combined\\\hline
      LEP Beam Energy            & 11 & 17 & 17 & 17 \\
      ISR/FSR                    &  7 &  8 &  9 &  8 \\
      Detector Systematics       &115 & 12 &  8 & 10 \\
      Hadronization              & -  & 19 & 17 & 17 \\
      Colour Reconnection        & -  &  - & 40 & 11 \\
      Bose-Einstein Correlations & -  &  - & 25 &  7 \\
      Other                      & 52 &  4 &  4 &  3 \\\hline
      Total Systematic           &127 & 29 & 54 & 30 \\\hline
      Statistical                &410 & 33 & 30 & 26 \\\hline\hline
      Total                      &430 & 44 & 62 & 40 \\\hline
    \end{tabular}
  \end{center}
  \label{tab:dmw}
\end{table}

\section{Colour Reconnection}

Since a theory that precisely describes the transition from coloured
partons, i.e.\ quarks and gluons, to colourless hadrons is currently
not available, one has to rely on phenomenological models of the
hadronization process. In the context of $W$ pair production and the
subsequent decay into four quarks, interactions between quarks from
different $W$'s are possible due to the smallness of the separation
between their decay vertices ({\cal O}$({0.1\,{\rm fm}})$) with
respect to typical hadronization lengths of {\cal O}(${1\,{\rm fm}}$).
Hence, there is no unique assignment of hadrons to $W$'s which may
systematically influence the direct reconstruction of the $W$ mass.
In terms of models that have been suggested (Sj"ostrand-Khoze or SK,
ARIADNE, HERWIG, Rathsman) the colour forces strung between the
partons may be reconnected if it leads to a more favourable
configuration according to a model-specific prescription.  Effects may
be observed e.g.\ for particle multiplicities, inter-jet particle
distributions and, unfortunately, for the $W$ mass itself.

\begin{figure}[t]
  \hftwo\raisebox{0.5cm}{\includegraphics[width=0.49\textwidth]{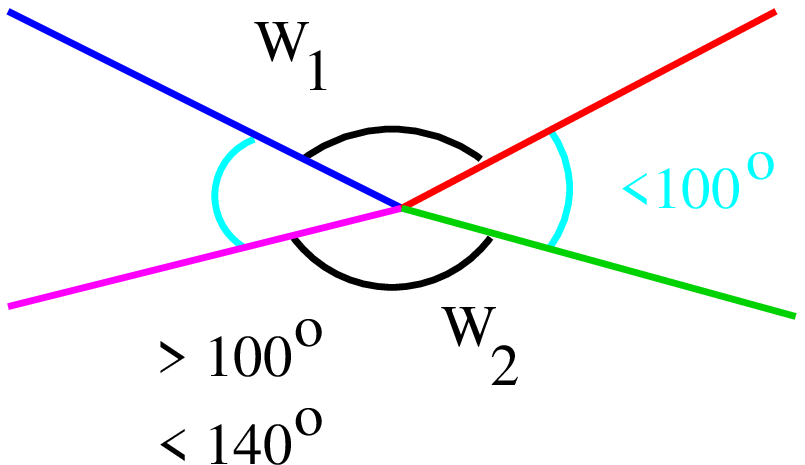}}\hftwo%
  \includegraphics[width=0.49\textwidth]{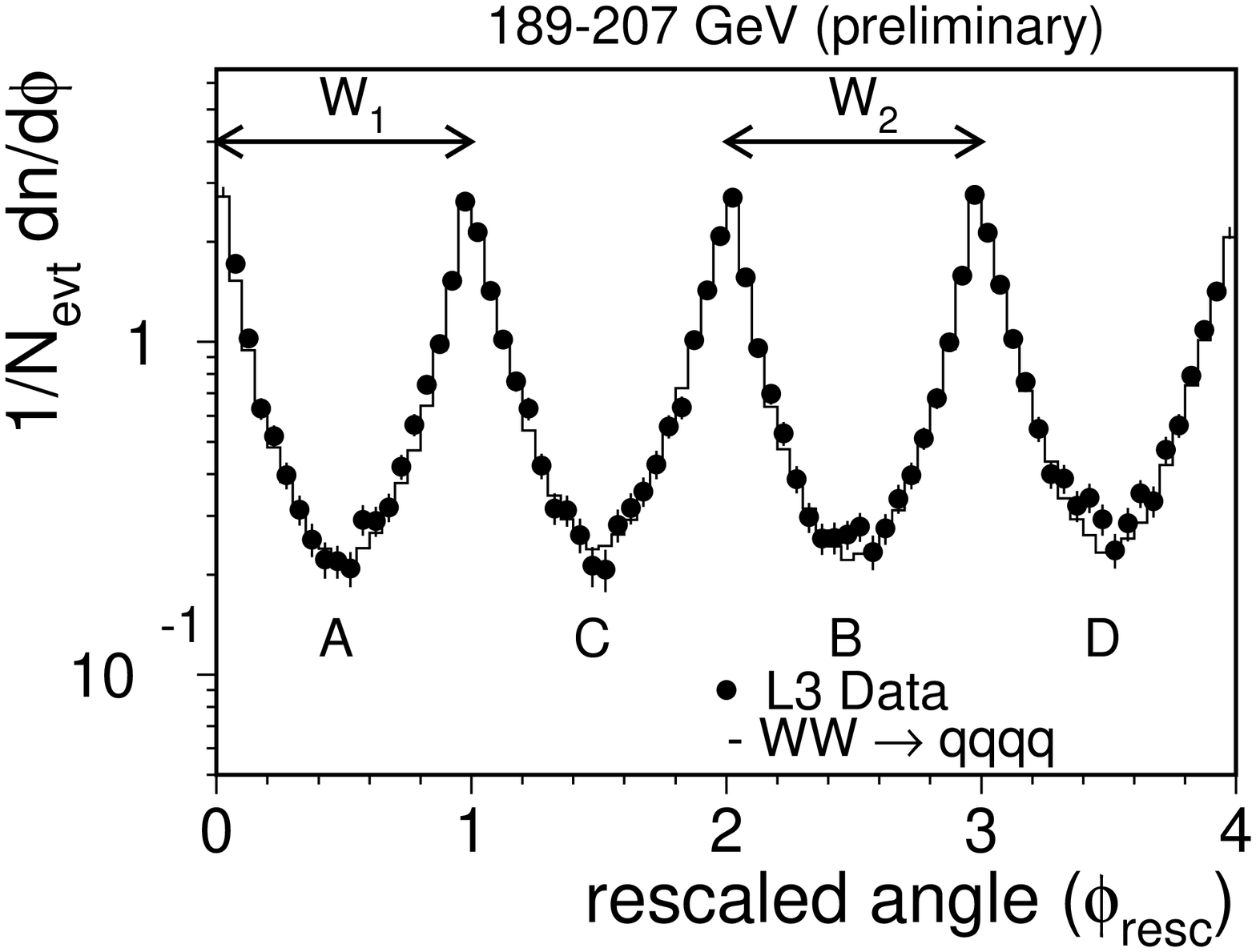}\hftwo
  \caption{A sketch of the additional requirements in the topological
    selection (left) and the particle flow with respect to the
    rescaled angle from \ldrei\ (right).}
  \label{fig:pflow}
  \vspace*{-0.40cm}
\end{figure}

Apart from the latter the inter-jet particle flow method, pioneered by
\ldrei,\cite{pflow} is one of the most sensitive so far. Compared to
the selection requirements for the $W$ mass analysis more stringent
cuts on the jet properties are imposed here (see fig.~\ref{fig:pflow}
left) leading to a cleaner jet topology at the expense of an
efficiency as low as $12\%$.  Currently, \delphi\ and \ldrei\ employ
this topological selection whereas \aleff\ and \opal\ prefer to stick
to their standard one.

Starting from the most energetic jet and continuing via the second jet
attributed to the same $W$ to the two jets of the second $W$, in total
four inter-jet planes are defined. Each particle is projected onto all
four planes and is assigned to the plane with which it makes the
smallest angle. It is included in the particle flow if the projection
lies between the two jets connected by the corresponding plane.  After
a rescaling of the angular distance to the jet defining the start of
the plane the particle flow can be shown as in the example from
\ldrei\ in fig.~\ref{fig:pflow} right.\cite{l3note} Effects of colour
reconnection can now be searched for in the form of differences
between the flow in intra-$W$ regions A and B and inter-$W$ regions C
and D with and without CR\@.

\begin{figure}[t]
  \hftwo\includegraphics[width=0.49\textwidth]{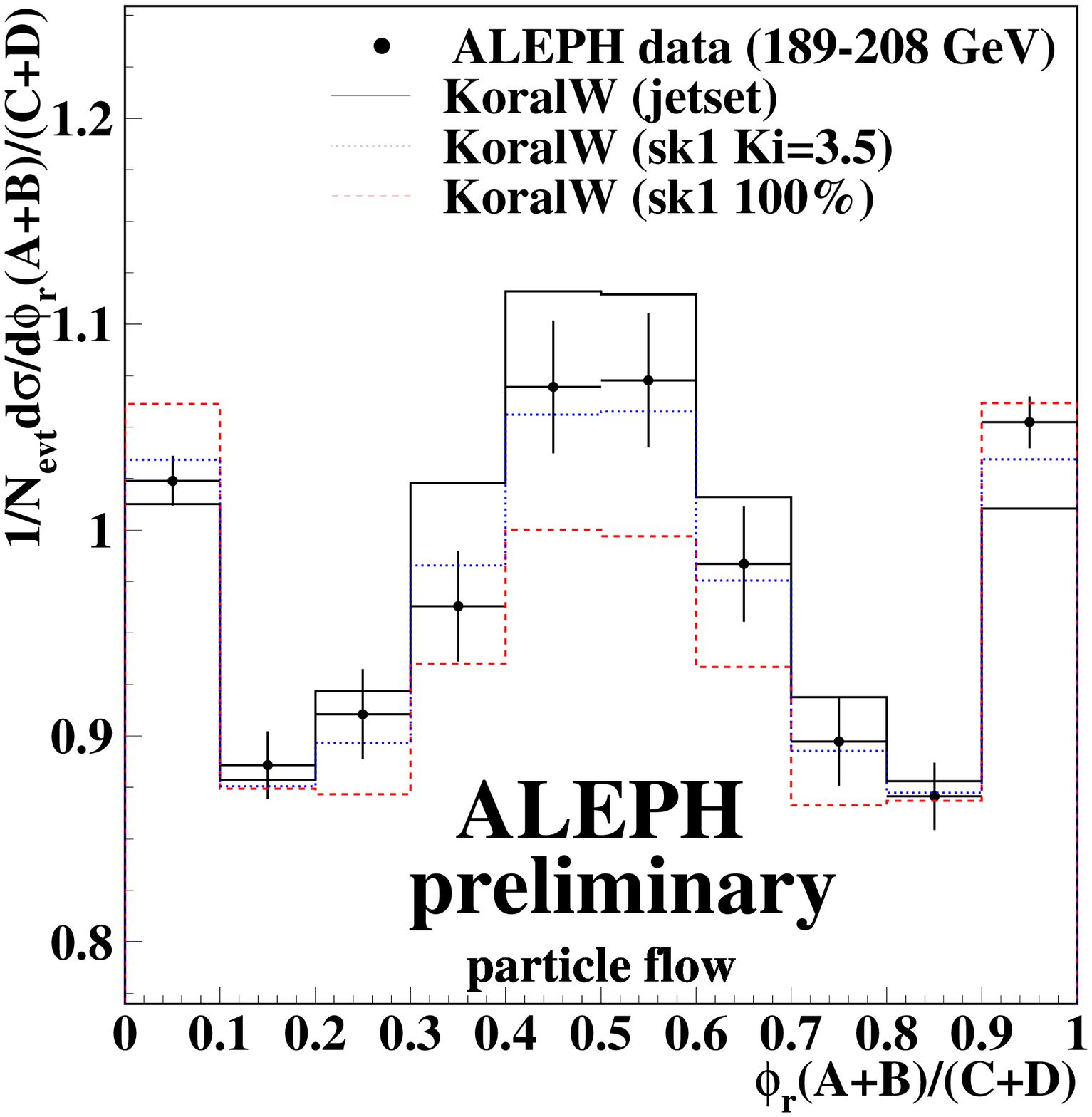}\hftwo%
  \begin{minipage}[b]{0.49\textwidth}
    \begin{center}
      \LARGE$R_N = \frac{\int_{0.2}^{0.8}
        \frac{1}{N}\cdot\frac{dn}{d\Phi_r}({\rm intra})d\Phi'_r}
      {\int_{0.2}^{0.8} \frac{1}{N}\cdot\frac{dn}{d\Phi_r}({\rm
          inter})d\Phi'_r}$ \normalsize
    \end{center}
    \vspace{0.5cm} \includegraphics[width=\textwidth]{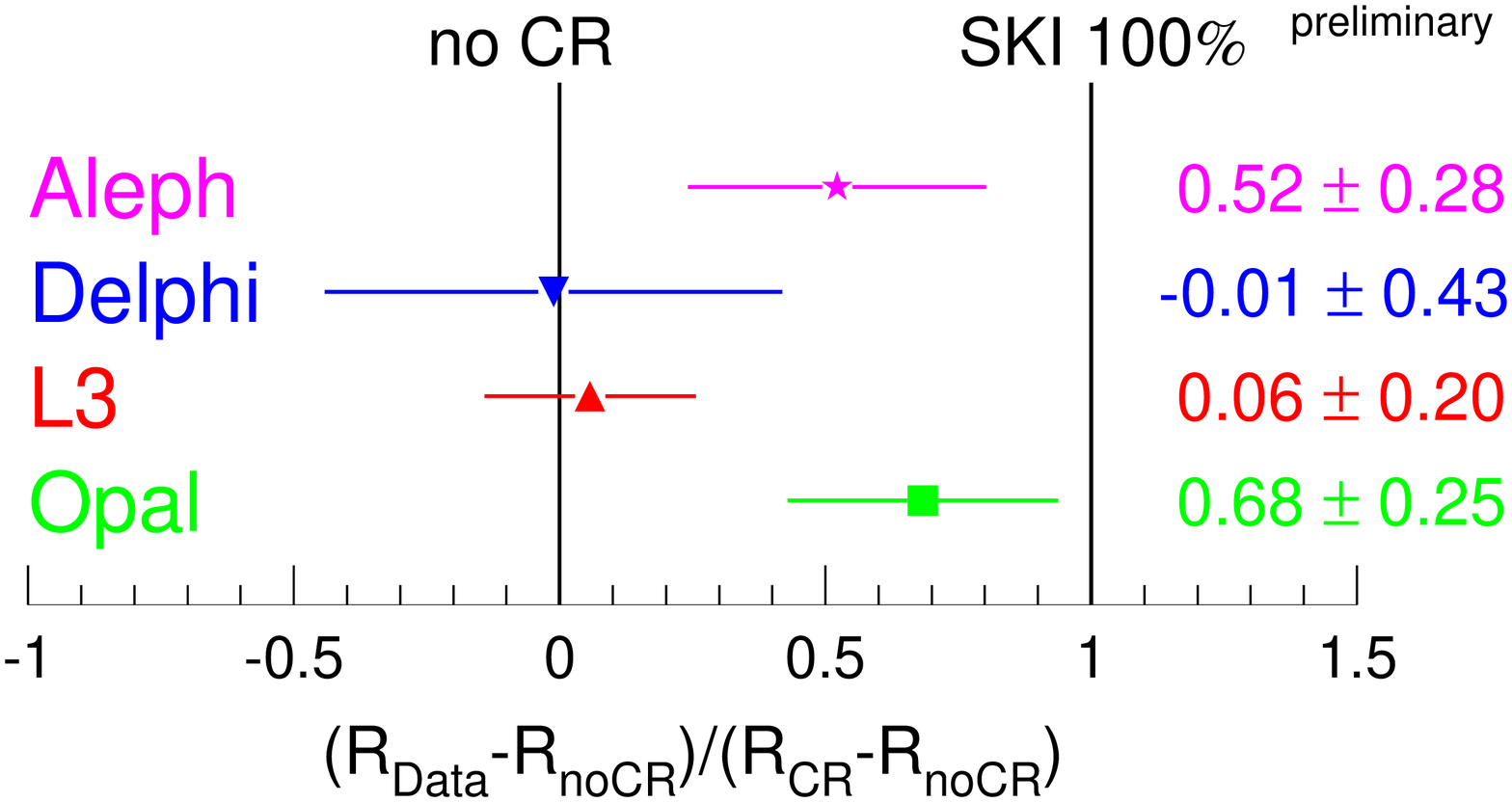}
    \vspace{0.5cm}
  \end{minipage}
  \rule{0.cm}{7.0cm}
  \caption{Ratio of intra- to inter-$W$ particle flow from \aleff\ (left),
    definition of integrated ratio $R_N$ and ratio of its offset with
    respect to a no CR scenario for data and MC (right).}
  \label{fig:rn}
  \vspace*{-0.25cm}
\end{figure}

Figure~\ref{fig:rn} left from \aleff~\cite{alephnote} shows the ratio
of the particle flow in the intra- and inter-$W$ regions for data and
three MC files with $0\%$ (full), $70\%$ (dotted) and $100\%$ (dashed)
reconnections for the SK I model. The extreme case of $100\%$
reconnections is disfavoured but values up to $80\%$ seem to be
consistent with the data. To focus on the region away from the
original jet direction the integral $R_N$ from $0.2$ up to $0.8$ is
used to extract the numbers in fig.~\ref{fig:rn} right.\cite{datong}
The more stringent selections employed by \delphi\ and \ldrei\ seem to
exhibit smaller reconnection effects but within current uncertainties
a clear conclusion can not be drawn.  A large LEP-wide effort is now
going into the production and simulation of common MC files for all
four experiments in order to address the problem of a consistent
combination of the individual results.

\vspace*{-0.20cm}
\section{Hadronization}
\vspace*{-0.25cm}

Table~\ref{tab:dmw} demonstrates that the modelling of hadronization
is a large uncertainty even without additional complications like
colour reconnection, and it is the dominant one that is in common for
the semi-leptonic and hadronic decay channels.  Individual estimates
by the experiments using the Jetset, Ariadne and Herwig MC's vary for
this uncertainty from $15\mev$ up to $30\mev$.  Also here the common
MC files will be used to address the somewhat incoherent picture,
which is partially due to different MC tunings used for the individual
experiments.

\vspace*{-0.20cm}
\section{Summary and Outlook}
\vspace*{-0.25cm}

The current preliminary LEP combined value for the mass of the $W$
boson is~\cite{lepcomb}:\\
\centerline{$M_W = (80.450 \pm 0.039)\gev$.}  A large part of the
total uncertainty is due to the necessity to use phenomenological
models for the hadronization process in general and for colour
reconnection in particular. Ongoing efforts in the LEP community aim
at a better understanding of the common problems in order to achieve
improved and consistent results that can be combined to reduce the $W$
mass uncertainty on the one hand but also to learn something about the
possible existence of final state interactions.  There is more to come
soon.

\end{document}